\documentclass[a4paper,12pt,fleqn]{article}

\usepackage[left=2.5cm,right=2.5cm,top=2.5cm,bottom=2.5cm]{geometry} 

\usepackage[latin1]{inputenc}
\usepackage{epsfig,graphics,graphicx,color}
\usepackage{amsmath,amssymb,amsthm,latexsym,mathrsfs}

\usepackage[pdftex,hyperfootnotes=false,pdfpagelabels,pdfstartview=FitH]{hyperref}
\usepackage[bf,font={singlespacing,small}]{caption}
\usepackage{subfig}
\usepackage{setspace}

\newcommand{\mailto}[1]{\href{mailto:#1}{#1}}

\title{\textbf{Measurement of the convective heat-transfer coefficient}}

\author{Rosaria Conti, Aurelio Agliolo Gallitto, Emilio Fiordilino \vspace{2mm} \\
\emph{Dipartimento di Fisica e Chimica, Università di Palermo} \\ \emph{via Archirafi 36, I-90123 Palermo, Italy}}

\date{}

\begin{document}

\maketitle

\begin{abstract}
\noindent We propose an experiment for investigating how objects cool down toward the thermal equilibrium with its surrounding through convection.
We describe the time dependence of the temperature difference of the cooling object and the environment with an exponential decay function.
By measuring the thermal constant $\tau$, we determine the convective heat-transfer coefficient, which is a characteristic constant of the convection system.

\end{abstract}

\vspace{12pt}

\begin{spacing}{1.2}
\noindent \textbf{PACS numbers} \vspace{6pt}\\
\begin{tabular}{ll}
  01.50.Pa & Laboratory experiments and apparatus\\
  44.25.+f & Natural convection\\
  01.30.la & Secondary schools
\end{tabular}


\vfill

\vfill

\noindent \textbf{Corresponding author} \vspace{6pt}\\
Prof. Aurelio Agliolo Gallitto\\
Dipartimento di Fisica e Chimica, University of Palermo\\
via Archirafi 36, I-90123 Palermo, Italy\\
Tel: +39 091 238.91702 -- Fax: +39 091 6162461\\
E-mail: \mailto{aurelio.agliologallitto@unipa.it}
\end{spacing}

\newpage
\section{Introduction}

In the last years, an increasing number of researchers are involved in programmes aimed at saving energy and reducing emission of carbon dioxide in the atmosphere.
In this context, the challenge to sustainability was issued by the U4energy Project \cite{U4energy}, an initiative funded through the Intelligent Energy Europe Programme of the European Union \cite{europa}.

Recently, scientific interest in the study of energy efficiency of houses has increased substantially since, in many countries, a large percentage of energy consumption is due to heating and cooling of buildings \cite{hewitt}.
For this reason, the research to improve the energy efficiency of buildings is a challenge for the next future.
The main process of energy exchange involved in home conditioning is convection.
Both natural and forced convection systems are employed in a common home heating system: e.g. hot water is circulated through radiators by forced convection, while warmed air rises by natural convection.

In this article we propose an experiment to measure the convective heat-transfer coefficient, which is a characteristic constant of convection systems.
This experiment can be also used as a laboratory introduction to exponential decay functions \cite{dewdney}.
The activity has been carried out together with students of secondary school, in the framework of the Italian National Plan for Scientific Degrees \cite{aracne}.

\section{Simplified model for cooling of objects}

Thermal energy is exchanged between two systems via conduction, convection and radiation \cite{blundell}.
In the case of cooling objects, from calorimetric relations one has that the heat released by the object is proportional to the temperature variation of the body assuming that the heat transfer is slow enough to maintain the temperature uniform within the object
\begin{equation}\label{eq:calore}
Q = - m c \Delta T \,,
\end{equation}
where $m$ is the mass and $c$ the specific heat of the object.\\
The energy flow $H$ can be written as
\begin{equation}\label{eq:flusso2}
H = \frac{dQ}{dt} = - m c \frac{d}{dt}T(t) \,,
\end{equation}
obtained by equation~\eqref{eq:calore} for $\Delta T \rightarrow 0$.
The negative sign indicates that the temperature decreases when the energy flows from the body toward outside (that is, when $H > 0$).

A comprehensive study of the heat transfer problem has been discussed by Vollmer \cite{vollmer}.
The author has shown that for convective heat-transfer coefficient $h_{conv} \approx 30~\mathrm{W~m^{-2}~K^{-1}}$ a simple exponential cooling seems to work quite well for $\Delta T < 100$~K.
In this case, the contribution to the energy exchange through radiation is about 10\% of convection and it becomes larger in the case of lower heat-transfer coefficients \cite{vollmer}.
Therefore, for $h_{conv} \gtrsim 30~\mathrm{W~m^{-2}~K^{-1}}$, small $\Delta T$ and for systems in which the emissivity coefficient $\varepsilon$ is smaller than the ideal black-body coefficient ($\varepsilon = 1$) the energy exchange by radiation can be neglected.

Following the simplified model for cooling of objects proposed by Vollmer, neglecting the contribution of radiation, one has
\begin{equation}\label{eq:equilibrio}
 m c \frac{d}{dt}T(t) = - h_{conv}~A~[T(t) - T_a] \,,
\end{equation}
where $A$ is the surface through which the heat flows, $T(t)$ the temperature of the external surface of the body at the time $t$ and $T_a$ the ambient temperature.\\
Equation \eqref{eq:equilibrio} can be rewritten as follows
\begin{equation}\label{eq:equilibrio2}
\frac{d}{dt}T(t) + \frac{T(t)}{\tau} = \frac{T_a}{\tau} \,,
\end{equation}
where we have introduced the thermal time constant $\tau$ as
\begin{equation}\label{eq:tau}
\tau = \frac{m~c}{h_{conv}~A} \,;
\end{equation}
$1/(h_{conv}A)$ is the ``thermal resistance'' and $mc$ the ``thermal capacitance'' \cite{thomsen}.
The time constant $\tau$ indicates that large masses $m$ and large specific heats $c$ (large thermal capacitances) give rise to slow temperature changes, whereas large surfaces $A$ and large thermal transfer coefficients $h_{conv}$ (small thermal resistances) give rise to fast temperature changes.
By supposing that $T_a$ is constant and replacing $\Delta T \equiv T(t) - T_a$, the solution of equation~\eqref{eq:equilibrio2} is
\begin{equation}\label{eq:delta_T}
\Delta T(t) = \Delta T_0 ~ e ^{-t/\tau}\,,
\end{equation}
where $\Delta T_0$ is the initial temperature difference, at $t=0$.\\
In other words, the object reaches the same temperature of the ambient following an exponential law determined by the characteristic thermal time constant $\tau$.
By measuring $\tau$, from equation~\eqref{eq:tau} we can determine the convective heat-transfer coefficient.

We would like to remark that the exponential dependence is valid for small temperature differences between the object and its surrounding.
In case of large temperature differences, different behaviors are observed \cite{vollmer,french}.
An analytical solution of cooling and warming laws has been recently proposed by Besson \cite{besson}.

\section{Experimental apparatus and results}
\begin{figure}[h!]
  \centering
  \includegraphics[width=0.65\textwidth]{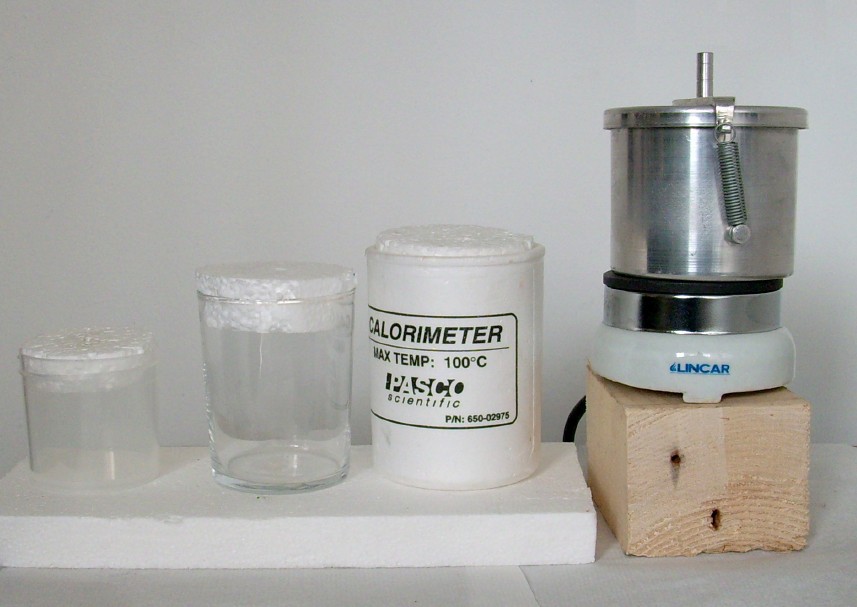}
  \caption{The containers used in the experiment: plastic (polypropylene), glass and expanded polystyrene shown from left to right. The heater used to warm up water is also shown on the right side.}\label{fig:contenitori}
\end{figure}

The experimental apparatus consists of three containers of different material and a computer equipped with an interface and a temperature sensor.
The interface used is the Vernier LabPro \cite{labpro}, which works under the Windows operative system and is controlled by the software Vernier LoggerPro; it allows one to automatically acquire experimental data by using sensors, as the Vernier temperature sensor.
The interface can acquire data with maximum sampling rate of $5 \times 10^4$ readings per second.
The Vernier temperature sensor is based on a NTC resistor, which has a resolution of about 0.1°C in the temperature range from 40°C to 100°C, accuracy of about $\pm 0.5$°C at $T \sim 100$°C and a response time of about 10~s, when immersed in water \cite{vernier}.
Three different containers made of glass, plastic (polypropylene) and expanded polystyrene have been used; they are shown in figure \ref{fig:contenitori}.

\begin{figure}[h!]
  \centering
  \includegraphics[width=0.65\textwidth]{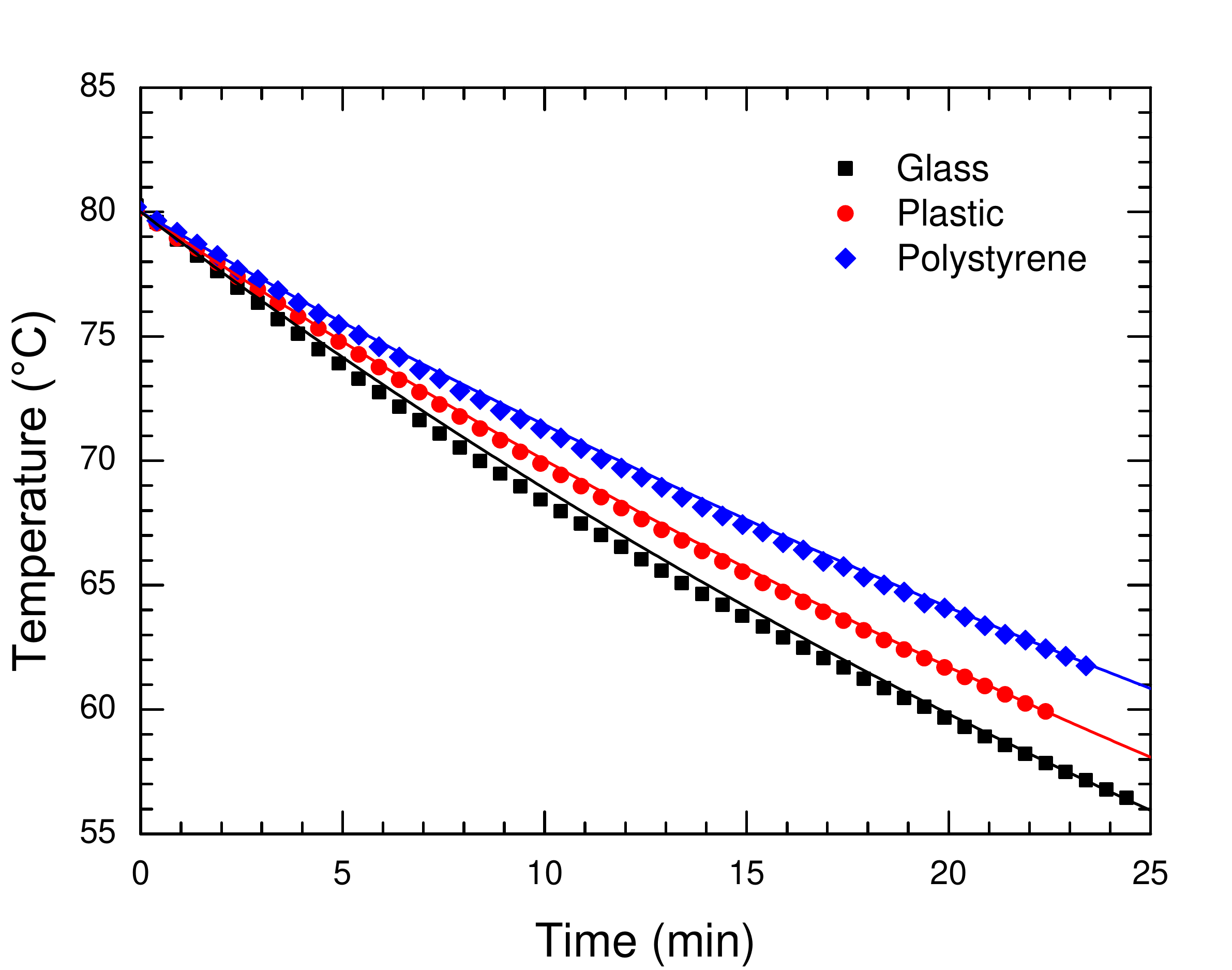}
  \caption{Temperature of the water as a function of time for the three different containers. Symbols are experimental data. Lines are theoretical curves obtained as explained in the text. The plots have been shifted in time such that they all start at the same temperature of 80°C.}\label{fig:dati}
\end{figure}

In order to measure the time dependence of the temperature during the cooling process, we have filled the containers with $m_w = 140~\mathrm{cm^3}$ of water at $T \approx 90$°C.
Soon after, we have measured the temperature of the water as a function of time for about 30~min.
The temperature values have been automatically acquired by the software LoggerPro.
In order to prevent that the energy flows from the top and from the bottom, the containers have been thermally insulated from the bottom and from the top by layers of expanded polystyrene with thickness of about 2~cm.
The cooling of the system has been done in still air under natural convection.
The results obtained with the different containers are reported in figure \ref{fig:dati}.

\section{Discussion}
The variation with time of the temperature of a liquid that cools in an environment at lower temperature can be described by an exponential law if the temperature difference between the object and the ambient is small.
An exponential law is peculiar of all the transitory phenomena that concern the evolution of the system toward an equilibrium state, or a stationary state.
Similar temporal dependencies can be observed in mechanical systems such as a falling body in a viscous medium, as well as in electric circuits such as the charging and discharging process of electric capacitors in a RC circuit, etc.
Since there is a common mathematical description for all these phenomena, it would be very fruitful for students to note that the same equations can describe experiments lying in different realms of the physics which, therefore, admit the same solution: in this case an exponential decay \cite{dewdney,vollmer,thomsen}.

By fitting the experimental data with equation~\eqref{eq:delta_T}, we have obtained the values of $\tau$.
As it can be seen in figure \ref{fig:dati}, the temperature variation follows an exponential decay quite well.
To determine the values of $h_{conv}$, we have to include also the mass $m_c$ and heat capacity $c_c$ of the containers used in the experiment.
From equation \eqref{eq:tau} we have
\begin{equation}\label{eq:h-conv}
h_{conv} = \frac{m_c~c_c + m_w~c_w}{\tau~A} \,.
\end{equation}
Since we need the lateral surface $A$ of the containers in contact with water through which the heat flows, we have determined the lateral surfaces by measuring the height $l$ of the water in the three containers and multiplying it by the external circumference of each container.
The values are shown in table \ref{tab:surface}.
\begin{table}[b]
\centering
\caption{Physical parameters of the materials used in the experiment; $D_e$ is the external diameter of the containers.}
\begin{tabular}{lccccc}
  \hline \hline
  \textbf{Material} & $m_c$ & $c_c$                 & $D_e$     & $l$         & A\\
                    &(g)    & (J g$^{-1}$ K$^{-1}$) & (mm)      & (mm)        & (cm$^2$)\\
  \hline
  Glass             & 152   & $0.50 - 0.84$         & $69\pm 1$ & $39 \pm 2$  & $84 \pm 4$\\
  Plastic           & 23.5  & 0.46                  & $65\pm 1$ & $44 \pm 2$  & $90 \pm 3$\\
  Polystyrene       & 8.2   & 1.3                   & $80\pm 2$ & $31 \pm 2$  & $78 \pm 5$\\
  Water             & 140   & 4.18                  &           &             &           \\
  \hline\hline
\end{tabular}
\label{tab:surface}
\end{table}

In table \ref{tab:tau}, we report the values of $\tau$ determined by best fitting the experimental data and $h_{conv}$ obtained by \eqref{eq:h-conv}, for each container.
In figure \ref{fig:dati}, we report the curves (lines) calculated by equation~\eqref{eq:delta_T}, using the estimated characteristic time $ \tau$ for each data set.
\begin{table}[t]
\centering
\caption{Values of the thermal parameters obtained from the fitting of the experimental data. The large uncertainty is due to the many factors that affects the results.}
\begin{tabular}{lcc}
  \hline \hline
  \textbf{Material} & $\tau$  & $h_{conv}$\\
                    & (min)   & $\mathrm{(W~m^{-2}~K^{-1})}$\\
  \hline
  Glass       & $49 \pm 2$     & $28 \pm 2$ \\
  Plastic     & $55 \pm 2$     & $21 \pm 2$ \\
  Polystyrene & $65 \pm 3$     & $19 \pm 2$ \\
  \hline\hline
\end{tabular}
\label{tab:tau}
\end{table}

We would remark that we have considered infinite the thermal conductivity of the containers, that is the temperature of the outer surface of the container is equal to the temperature of the inner surface, and supposed that the temperature of the cooling liquid was uniform in each point and on its surface at any time and that it uniformly changes with time \cite{lewis}.
Furthermore, the proposed model does not take into account energy exchange through radiation; therefore, the values of the heat-transfer coefficients we have determined could be affected by an uncertainty larger than that indicated in table~\ref{tab:tau}, especially for plastic and polystyrene containers since for them $h_{conv} < 30~\mathrm{W~m^{-2}~K^{-1}}$.

\section{Conclusion}
We have performed an experiment for investigating how objects cool down toward the thermal equilibrium with its surrounding.
We have described the time dependence of the temperature difference of the cooling objects and the environment with an exponential decay function.
By measuring the thermal constant $\tau$, we have determined the convective heat-transfer coefficient characteristic of the system.
As expected, water in the container made of expanded polystyrene cools down at the slowest rate, whereas in the container made of glass cools down at the fastest rate.
The proposed model can be applied for the experimental determinations of such coefficients using different containers under various experimental conditions.
This activity can be also used as a laboratory introduction to exponential decay functions for high-school students.
Although the simplified model is valid for convective heat-transfer coefficient of about 30~$\mathrm{W~m^{-2}~K^{-1}}$ and for $\Delta T < 100$~K, that is when the energy exchange by radiation is negligible, it could be easily performed at schools allowing teachers to discuss energy exchange processes and it might contribute to increase the awareness of students toward  energy saving and sustainability issues.

\section*{Acknowledgements}
\addcontentsline{toc}{section}{Acknowledgements}
This work was carried out under the financial support of the Italian Ministry of Education, University and Research.

\end{document}